\documentclass[aps,prl,twocolumn,groupedaddress]{revtex4-1}
\usepackage{graphicx}% Include figure files
\usepackage{amsmath}
\usepackage{dcolumn}% Align table columns on decimal point
\usepackage{bm}% bold math

\begin{document}

\preprint{}

\title{Exciton Diamagnetic Shifts and Valley Zeeman Effects in Monolayer WS$_2$ and MoS$_2$ to 65 Tesla}

\author{Andreas V. Stier$^1$, Kathleen M. McCreary$^2$, Berend T. Jonker$^2$, Junichiro Kono$^3$, Scott A. Crooker$^1$}

\affiliation{$^1$National High Magnetic Field Laboratory, Los Alamos, New Mexico 87545, USA}
\affiliation{$^2$Materials Science and Technology Division, Naval Research Laboratory, Washington, DC 20375, USA}
\affiliation{$^3$Department of Electrical and Computer Engineering, Department of Physics and Astronomy, and Department of Materials Science and NanoEngineering, Rice University, Houston, Texas 77005, USA}

\begin{abstract}
We report circularly-polarized optical reflection spectroscopy of monolayer WS$_2$ and MoS$_2$ at low temperatures (4~K) and in high magnetic fields to 65~T.  Both the A and the B exciton transitions exhibit a clear and very similar Zeeman splitting of approximately $-$230~$\mu$eV/T ($g\simeq -4$), providing the first measurements of the valley Zeeman effect and associated $g$-factors in monolayer transition-metal disulphides. These results complement and are compared with recent low-field photoluminescence measurements of valley degeneracy breaking in the monolayer diselenides MoSe$_2$ and WSe$_2$. Further, the very large magnetic fields used in our studies allows us to observe the small quadratic diamagnetic shifts of the A and B excitons in monolayer WS$_2$ (0.32 and 0.11~$\mu$eV/T$^2$, respectively), from which we calculate exciton radii of $\sim$1.53~nm and $\sim$1.16~nm. When analyzed within a model of non-local dielectric screening in monolayer semiconductors, these diamagnetic shifts also constrain and provide estimates of the exciton binding energies ($\sim$410~meV and $\sim$470~meV for the A and B excitons, respectively), further highlighting the utility of high magnetic fields for understanding new 2D materials.
\end{abstract}

\maketitle
Atomically-thin crystals of the transition-metal disulphides (MoS$_2$ and WS$_2$) and diselenides (MoSe$_2$ and WSe$_2$) constitute a novel class of monolayer semiconductors that possess direct optical bandgaps located at the degenerate $K$ and $K'$ valleys of their hexagonal Brillouin zones \cite{Mak2010, Splendiani}. The considerable recent interest in these 2D transition-metal dichalcogenides (TMDs) derives from their strong spin-orbit coupling and lack of structural inversion symmetry which, together with time-reversal symmetry, couples spin and valley degrees of freedom and leads to valley-specific optical selection rules \cite{Xiao, MakNatNano, Zeng, Sallen, Cao, XuReview}: light of $\sigma^+$ circular polarization couples to inter-band exciton transitions in the $K$ valley, while the opposite ($\sigma^-$) circular polarization couples to transitions in the $K'$ valley. The ability to populate and/or probe electrons and holes in specific valleys using polarized light has renewed long-standing interests \cite{XuReview, Sham, Shayegan, graphene} in understanding and exploiting such `valley pseudospin' degrees of freedom for both fundamental physics and far-reaching applications in, \emph{e.g.}, quantum information processing.

The bands and optical transitions at the $K$ and $K'$ valleys are nominally degenerate in energy and related by time-reversal symmetry. However, in analogy with conventional spin degrees of freedom, this $K/K'$ valley degeneracy can be lifted by external magnetic fields $B$, which break time-reversal symmetry. Recent photoluminescence (PL) studies of the monolayer diselenides MoSe$_2$ and WSe$_2$ in modest fields have indeed demonstrated this `valley Zeeman effect', and revealed an energy splitting between $\sigma^+$ and $\sigma^-$ polarized PL from the lowest-energy ``A" exciton transition \cite{Li, MacNeill, Aivazian, Srivastava, Wang, Mitioglu}. In most cases, valley splittings in these monolayer diselenides were found to increase linearly with field at a rate of approximately $-$4$\mu_{\rm B}$ ($\equiv -232$~$\mu$eV/T), where $\mu_{\rm B}=57.9$~$\mu$eV/T is the Bohr magneton. While this value agrees surprisingly well with simple expectations from a two-band tight-binding model (namely, that electron and hole masses are equal, and that the exciton Zeeman shifts derive solely from the hybridized $d_{x^2-y^2} \pm id_{xy}$ atomic orbitals with magnetic moment $\pm 2 \mu_\textrm{B}$ that comprise the $K/K'$ valence bands \cite{MacNeill, Li, Aivazian, Srivastava}), it is also widely appreciated that a more complete treatment based on established $\mathbf{k} \cdot \mathbf{p}$ theory should, with proper inclusion of strong excitonic effects, also provide an accurate description.  However, initial $\mathbf{k} \cdot \mathbf{p}$ models have so far had limited success accounting for the measured valley Zeeman effect in monolayer TMDs \cite{MacNeill, KormanyosPRX, Wang}.

At all events, magneto-optical studies of these new monolayer semiconductors are still at a relatively early stage and several outstanding questions remain. In particular, measurements of valley Zeeman effects in the monolayer di\emph{sulphides} WS$_2$ and MoS$_2$ have not been reported to date, which would provide a natural complement to the existing data on monolayer WSe$_2$ and MoSe$_2$. In addition, the valley Zeeman splitting of the higher-energy ``B" exciton has not yet been reported in any of these 2D materials. Both of these studies would provide a more complete experimental basis against which to benchmark new theoretical approaches. And finally, the \emph{diamagnetic} energy shift of these excitons, which is anticipated to increase quadratically with field and from which the spatial extent of the fundamental (1$s$) exciton wavefunctions can be directly inferred \cite{Knox, Miura, Walck}, has not yet been observed in any of the monolayer TMDs. Likely this is because the diamagnetic shift, $\Delta E_{\rm dia} = e^2 \langle r^2 \rangle_{1s} B^2/8 m_r$, is expected to be very small and difficult to spectrally resolve in these materials owing to the small root-mean-square (rms) radius of the 1$s$ exciton ($r_1 = \sqrt{\langle r^2 \rangle_{1s}}$), and large reduced mass $m_r =(m_e^{-1} + m_h^{-1})^{-1}$. For example, if $r_1 \approx 1.5$~nm and $m_e = m_h \approx m_0/2$ (where $m_0$ is the bare electron mass and $m_{e/h}$ is the effective electron/hole mass), then $\Delta E_{\rm dia}$ is only $\sim$20~$\mu$eV at $B=10$~T, clearly motivating the need for large magnetic fields. Crucially, knowledge of $\Delta E_{\rm dia}$ can also constrain estimates of the exciton binding energy -- a subject of considerable recent interest in the monolayer TMDs \cite{Berkelbach, Chernikov, Ye, Zhu, He, Hanbicki, WangPRL2015, Stroucken, Ram, Shi, Komsa, Ugeda, Kirill} wherein the effects of non-local dielectric screening and Berry curvature can generate a markedly non-hydrogenic Rydberg series of exciton states and associated binding energies \cite{Keldysh, Cudazzo, Berkelbach2, Srivastava2, Zhou2}.

To address these questions, we report here a systematic study of circularly-polarized magneto-reflection from large-area films of monolayer WS$_2$ and MoS$_2$ at low temperatures (4~K) and in very high pulsed magnetic fields up to 65~T. Clear valley splittings of about $-$230~$\mu$eV/T are observed for \emph{both} the A and B excitons, providing the first measurements of the valley Zeeman effect and associated $g$-factors in monolayer transition-metal disulphides. Moreover, due to the very large magnetic fields used in these studies, we are also able to resolve the small quadratic diamagnetic shifts of both A and B excitons in monolayer WS$_2$ ($0.32 \pm 0.02~\mu$eV/T$^2$ and $0.11 \pm 0.02~\mu$eV/T$^2$, respectively), permitting estimates of the rms exciton radius $r_1$. These results are compared with similar measurements of bulk WS$_2$ crystals, and are quantitatively modeled within the context of the non-hydrogenic binding potential \cite{Keldysh, Cudazzo, Berkelbach} that is believed to exist in 2D semiconductors due to non-local dielectric screening.  Within this framework, we estimate A and B exciton binding energies of $\sim$410~meV and $\sim$470~meV, respectively, and we show how these values scale with reduced mass $m_r$.

\textbf{Experiment.} Large-area samples of monolayer WS$_2$ and MoS$_2$ were grown by chemical vapor deposition on SiO$_2$/Si wafers~\cite{McCreary}. MoO$_3$ and pure sulfur powder were used as precursor and reactant materials, and the growth was performed at a reactant temperature of 625~$^\circ$C. In addition, perylene-3,4,9,10-tetracarboxylic acid tetrapotassium salt was loaded on the SiO$_2$/Si substrate, which acted as a seeding promoter to achieve uniform large-area monolayer crystals \cite{Ling}. In addition, a freshly-exfoliated surface of a bulk WS$_2$ crystal was also prepared.

Magneto-reflectance studies were performed at cryogenic temperatures (down to 4~K) in a capacitor-driven 65~T pulsed magnet at the National High Magnetic Field Laboratory in Los Alamos. Broadband white light from a xenon lamp was coupled to the samples using a 100~$\mu$m diameter multimode optical fiber. The light was focused onto the sample at near-normal incidence using a single aspheric lens, and the reflected light was refocused by the lens into a 600~$\mu$m diameter collection fiber. A thin-film circular polarizer mounted over the delivery or collection fiber provided $\sigma^+$ or $\sigma^-$ circular polarization sensitivity. The collected light was dispersed in a 300~mm spectrometer and detected with a charge-coupled device (CCD) detector. Spectra were acquired continuously at a rate of 500~Hz throughout the $\sim$50~ms long magnet pulse.

\begin{figure}[tbp]
\center
\includegraphics[width=.45\textwidth]{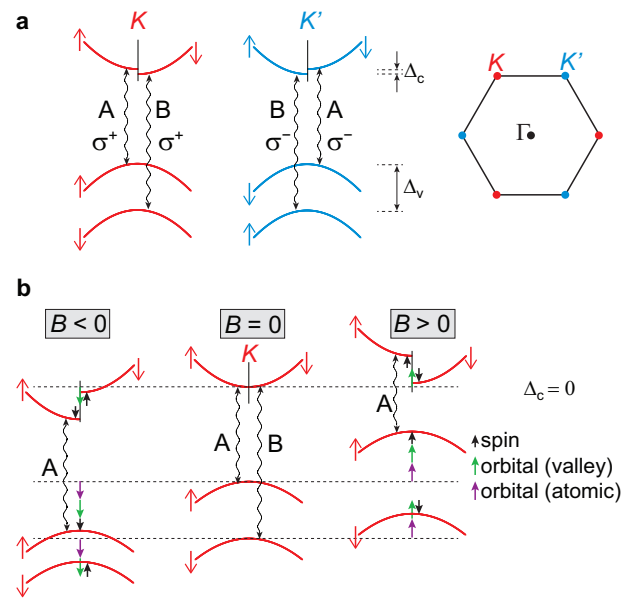}
\caption{\textbf{Excitonic transitions and Zeeman shifts in monolayer TMDs.} (a) Diagrams of the conduction and valence bands at the $K$ and $K'$ valleys of the monolayer transition-metal dichalcogenides, showing the A and B exciton transitions (wavy lines) and the associated valley-specific optical selection rules for $\sigma^+$ and $\sigma^-$ light. For clarity, the spin-up and spin-down conduction bands are separately drawn on the left and right side within each valley, respectively. A small spin-orbit splitting in the conduction band, $\Delta_\textrm{c}$, is also depicted for completeness (theory predicts $\Delta_\textrm{c} \sim$~+30 and $-$5 meV for WS$_2$ and MoS$_2$, respectively \cite{Liu}). (b) Diagrams depict the relative shifts of the conduction and valence bands in the $K$ valley (\emph{i.e.}, $\sigma^+$ transitions) due to applied magnetic fields $\pm B \parallel \hat{z}$. For clarity, $\Delta_\textrm{c}=0$ here. The contributions to each band's Zeeman shift from spin, atomic orbital, and valley orbital (Berry curvature) magnetic moment are depicted separately by black, purple, and green arrows. The $\sigma^+$ polarized A and B exciton transition energies decrease (increase) in positive (negative) field. By time-reversal symmetry, the shifts depicted here for $B<0$ in the $K$ valley are equivalent to those in the $K'$ valley ($\sigma^-$ transitions) when $B>0$.} \label{fig1}
\end{figure}

\textbf{Exciton transitions and Zeeman effects in monolayer TMDs.} Figure 1a depicts a single-particle energy diagram of the conduction and valence bands in monolayer TMDs at the $K$ and $K'$ points of the hexagonal Brillouin zone, along with the A and B exciton transitions (wavy lines) and valley-specific optical selection rules. Strong spin-orbit coupling of the valence band splits the spin-up and spin-down components (by $\Delta_\textrm{v} \sim$400~meV and 150~meV in WS$_2$ and MoS$_2$, respectively), giving rise to the well-separated A and B exciton transitions that are observed in optical absorption or reflection spectra.  As depicted, $\sigma^+$ circularly-polarized light couples to both the A and B exciton transitions in the $K$ valley, while light of the opposite $\sigma^-$ circular polarization couples to the exciton transitions in the $K'$ valley.

At zero magnetic field, time-reversed pairs of states in the $K$ and $K'$ valleys -- \emph{e.g.}, spin-up conduction (valence) bands in $K$ and spin-down conduction (valence) bands in $K'$ -- necessarily have the same energy and have equal-but-opposite total magnetic moment ($\bm{\mu}_K^{\textrm{c,v}} = - \bm{\mu}_{K'}^{\textrm{c,v}}$). Therefore, an applied magnetic field, which breaks time-reversal symmetry, will lift the $K/K'$ valley degeneracy by shifting time-reversed pairs of states in opposite directions in accord with the Zeeman energy $-\bm{\mu} \cdot \textbf{B}$. This will Zeeman-shift the measured exciton energy if the relevant conduction and valence band moments are unequal; \emph{viz}, $\Delta E_\textrm{Z} = -(\bm{\mu}^\textrm{c} - \bm\mu^\textrm{v})\cdot \textbf{B}$. In the following, we consider strictly out-of-plane fields, $\textbf{B}=\pm B \hat{z}$.

Figure 1b depicts the field-dependent energy shifts of the conduction and valence bands in the $K$ valley ($\sigma^+$ polarized light), for both positive and negative field. The various contributions to the total Zeeman shift in the monolayer TMDs have been discussed in several recent reports \cite{Xiao, Li, MacNeill, Aivazian, Srivastava, Yao}, which we summarize as follows. In general, the total magnetic moment $\bm{\mu}$ of any given conduction or valence band in the $K$ or $K'$ valley contains contributions from three sources: spin ($\mu_s$), atomic orbital ($\mu_l$), and the valley orbital magnetic moment that is associated with the Berry curvature ($\mu_k$). Note that the latter two have been referred to as ``intra-cellular" and ``inter-cellular" contributions to the orbital magnetic moment, respectively \cite{MacNeill, Srivastava}. The spin contribution to the exciton Zeeman shift $\Delta E_\textrm{Z}$ is expected to be zero, since the optically-allowed transitions couple conduction and valence bands having the same spin ($\mu_s^\textrm{c} = \mu_s^\textrm{v}$). In contrast, the conduction and valence bands are comprised of entirely different atomic orbitals: the $d_{z^2}$ orbitals of the conduction bands have azimuthal orbital angular momentum $l_z=0$ ($\mu_l^{\rm c}=0$), while the hybridized $d_{x^2-y^2} \pm id_{xy}$ orbitals that comprise the valence bands have $l_z=\pm 2 \hbar$ ($\mu_l^{\rm v} = \pm 2 \mu_{\rm B}$) in the $K$ and $K'$ valleys, respectively. This contribution is expected to generate a Zeeman shift of the $K$ and $K'$ exciton of $\mp 2 \mu_\textrm{B} B$, respectively, and therefore, a total exciton \emph{splitting} of $-4 \mu_{\rm B} B$.  Finally, the valley orbital (Berry curvature) contribution to the conduction and valence band moments are $\mu_k^\textrm{c} = \pm (m_0/m_e)\mu_{\rm B}$ and $\mu_k^\textrm{v} = \pm (m_0/m_h)\mu_{\rm B}$ in the $K$ and $K'$ valleys, respectively. In a simple two-band tight-binding model where $m_e = m_h$, then $\mu_k^\textrm{c} = \mu_k^\textrm{v}$ and shifts due to the valley orbital magnetic moment do not appear in $\Delta E_\textrm{Z}$. In more general models \cite{Liu} where $m_e \neq m_h$, these Berry curvature contributions may play a role and cause a deviation of the exciton Zeeman splitting away from $-4 \mu_\textrm{B}$.

\begin{figure}[tbp]
\center
\includegraphics[width=.46\textwidth]{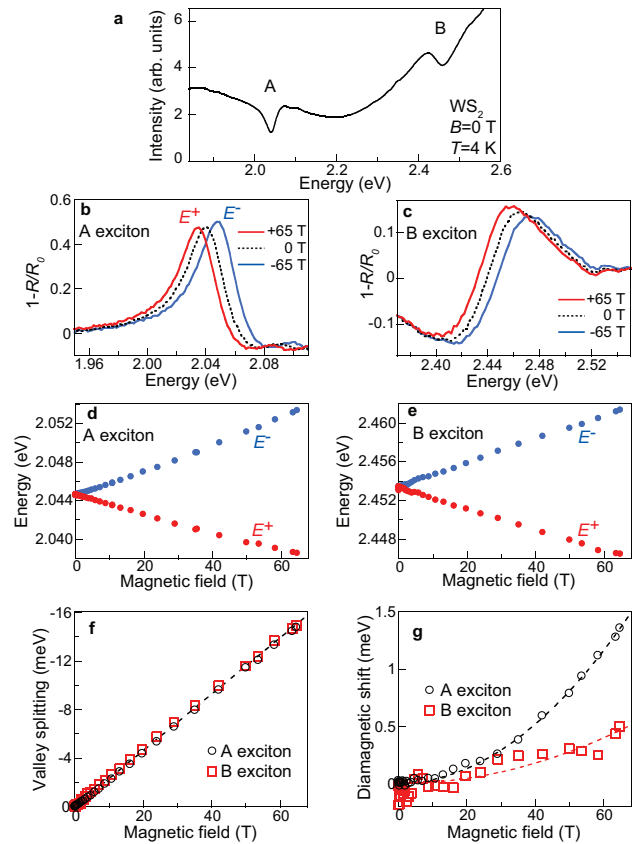}
\caption{\textbf{Valley Zeeman effect and diamagnetic shift in monolayer WS$_2$.} (a) Reflection spectrum of monolayer WS$_2$ at $B=0$~T and $T=4$~K. The A and B exciton resonances are labeled. (b) Normalized reflection spectra ($1-R/R_0$) at the A exciton resonance using $\sigma^+$ polarized light. The dashed black trace was acquired at $B=0$~T. The red trace was acquired at +65~T, and corresponds to the $\sigma^+$ transition in the $K$ valley. The blue trace was acquired at $-$65~T, which is equivalent (by time-reversal symmetry) to the $\sigma^-$ transition in the $K'$ valley at +65~T. The valley Zeeman splitting between these two peaks is clearly resolved. (c) Similar reflection spectra and valley Zeeman splitting of the B exciton. (d) Energies ($E^+$ and $E^-$) of the field-split A exciton versus magnetic field. (e) Same, but for the B exciton. (f) The measured valley Zeeman splitting ($E^+ - E^-$) versus magnetic field, for both A and B excitons. The dotted line corresponds to a splitting of $-4\mu_{\rm B}$ ($-$232 $\mu$eV/T). (g) The \emph{average} energy, ($E^+ + E^-$)/2, for both the A and B excitons (the zero-field offset has been subtracted). A small quadratic diamagnetic shift is revealed. The dotted lines show quadratic fits to the data ($\Delta E_{\rm dia} = \sigma B^2$), where $\sigma$ is the diamagnetic shift coefficient.  We find that $\sigma_\textrm{A}=0.32 \pm 0.02~\mu$eV/T$^2$ and $\sigma_\textrm{B}=0.11 \pm 0.02~\mu$eV/T$^2$ for the A and B exciton, respectively.} \label{fig2}
\end{figure}

To selectively probe the $K$ and $K'$ transitions in our magneto-reflectivity experiments, we typically fixed the handedness of the circular polarizer to $\sigma^+$, and pulsed the magnet in the positive (+65~T) and then the negative ($-$65~T) field direction. The latter case is exactly equivalent (by time-reversal symmetry) to measuring the $\sigma^-$ optical transitions in positive field (we also explicitly verified this by changing the circular polarizer). Sign conventions were confirmed via magneto-reflectance from a diluted magnetic semiconductor (Zn$_{0.92}$Mn$_{0.08}$Se) \cite{Furdyna}.

\textbf{Valley Zeeman effect in monolayer WS$_2$.} Figure 2a shows the reflection spectrum (raw data) from monolayer WS$_2$ at 4~K. Both the A and B exciton transitions are clearly visible and are superimposed on a smoothly-varying background. Figure 2b shows the well-resolved Zeeman splitting of the A exciton in WS$_2$ at the maximum $\pm65$~T applied magnetic field. Red, blue, and (dashed) black curves show the normalized reflection spectra, $1-R/R_0$ (where $R_0$ is a smooth background), at +65, $-$65, and 0~T respectively. A valley splitting of $\sim$15~meV, analyzed in detail below, is observed. Moreover, because these measurements are based on magneto-reflectance spectroscopy (rather than PL), the valley splitting of the higher-energy B exciton in WS$_2$ can also be observed, as shown in Fig. 2c.  For both the A and B exciton, the energy of the exciton transition in positive magnetic fields (hereinafter called $E^+$) shifts to lower energy, while the exciton energy in negative fields ($E^-$) shifts to higher energy, as labeled.

The exciton resonances were fit using complex (absorptive + dispersive) Lorentzian lineshapes to extract the transition energy. The field-dependent energies of the split peaks in monolayer WS$_2$, $E^+(B)$ and $E^-(B)$, are shown in Figs.~2d and 2e for the A and B excitons, respectively. The \emph{splitting} between the two valleys, $E^+ - E^-$, is shown in Fig.~2f for both the A and B excitons. The measured valley Zeeman splitting is negative, and increases in magnitude linearly with applied field, with nearly identical rates of $-228 \pm 2~\mu$eV/T for the A exciton and $-231 \pm 2~\mu$eV/T for the B exciton. These values correspond to Land\'{e} $g$-factors of $-3.94 \pm 0.04$ and $-3.99 \pm 0.04$, respectively, and represent the first measurements of the valley Zeeman effect in the monolayer transition-metal disulphides, and also the first measurement of the B exciton valley splitting in any monolayer TMD material.

\begin{figure}[tbp]
\center
\includegraphics[width=.47\textwidth]{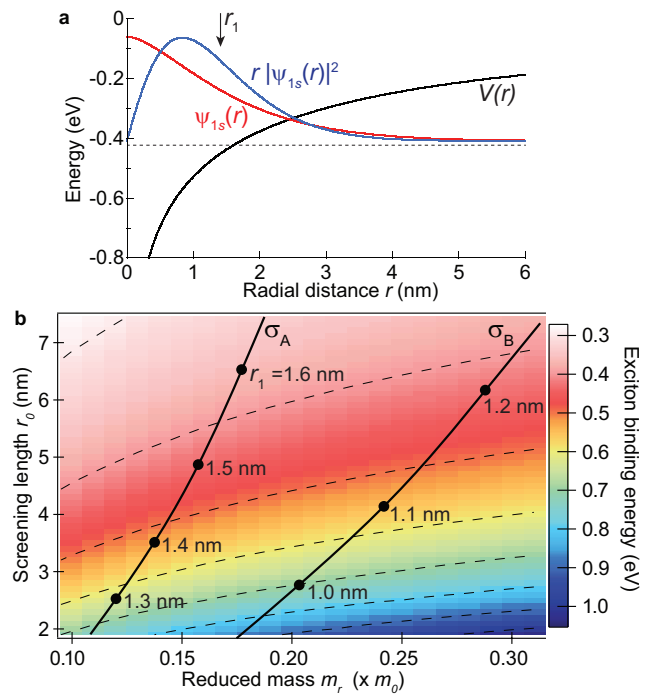}
\caption{\textbf{Constraining exciton binding energies via the diamagnetic shift.} (a) A plot of the (non-hydrogenic) 1$s$ A exciton wavefunction in monolayer WS$_2$, $\psi_{1s}(r)$, computed by numerically solving Schr\"{o}dinger's equation using a reduced mass $m_r=0.16$ and the non-local dielectric screening potential $V(r)$ defined in Eq.~(1). The screening length $r_0$ was adjusted to give $\psi_{1s}(r)$ such that the rms exciton radius $r_1 = \sqrt{\langle \psi_{1s}| r^2 |\psi_{1s} \rangle} = 1.53$~nm, which is the value calculated from the measured diamagnetic shift $\sigma_\textrm{A} = 0.32~\mu$eV/T$^2$. [Note that $r_1$ is an rms value, and does not correspond to the peak of the 2D radial probability density $r|\psi(r)|^2$.] (b) Color surface plot of the calculated exciton binding energy, using $V(r)$ from Eq.~(1), over a range of reduced mass $m_r$ and screening length $r_0$. Dashed lines show contours of constant binding energy. Solid lines indicate contours of constant diamagnetic shift corresponding to those measured in Fig.~2g for the A and B excitons in monolayer WS$_2$. The calculated rms exciton radius, $r_1$, is indicated at various points along the contours.} \label{fig3}
\end{figure}

The A exciton valley splitting that we measure in monolayer WS$_2$ is quite close to that reported recently from magneto-PL studies of its diselenide counterpart, monolayer WSe$_2$ \cite{Srivastava, Wang}. For comparison, reported $g$-factors for all the monolayer TMDs are shown in Table~1. As discussed above, our measured values of $g \simeq -4$ agree surprisingly well with a simple two-band tight-binding model, wherein $m_e = m_h$ and valley moment (Berry curvature) contributions to the exciton magnetic moment cancel out, so that the exciton Zeeman shifts derive solely from atomic orbital magnetic moments of the valence bands. However, Berry curvature contributions to the Zeeman splitting \emph{are} expected in more general models \cite{Liu} where $m_e \neq m_h$. Deviations away from $g=-4$, observed for example in \cite{Srivastava} and \cite{Aivazian}, have been explained along these lines (although, note that for tightly-bound excitons, the total valley moment contribution can vary significantly in magnitude and sign, because this orbital moment must be averaged over a substantial portion of the Brillouin zone \cite{Srivastava}).

In view of the above, it is therefore particularly noteworthy that we \emph{also} measure $g \simeq -4$ for the B exciton in monolayer WS$_2$, despite the fact that its reduced mass almost certainly differs from that of the A exciton, as shown below from direct measurements of the diamagnetic shift (\emph{i.e.}, $m_h$ cannot equal $m_e$ for both spin-up \emph{and} spin-down valence bands). Note that early studies of bulk MoS$_2$ \cite{Evans1967, Evans1968, Neville} also indicate that the B exciton mass significantly exceeds that of the A exciton. This suggests that contributions to the orbital moment from Berry curvature effects, expected when $m_e \neq m_h$, may not play a significant role in determining the measured exciton magnetic moment and the valley Zeeman effect.

\begin{table*}
\centering
\begin{tabular}{l l l l l}
\hline\hline
Material & A exciton \emph{g}-factor~~~~~~~~~~ & B exciton \emph{g}-factor~~~~~~ & $\sigma_\textrm{A}$ [$\mu$eV/T$^{2}$] & $\sigma_\textrm{B}$ [$\mu$eV/T$^{2}$]\\[0.5ex]
\hline
monolayer WS$_{2}$ & \begin{tabular}{@{}l@{}}\textbf{-3.94}~$\pm$~\textbf{0.04}\\\textbf{-3.33}~$\pm$~\textbf{0.04} (bulk) \\ -3.2~$\pm ~ 0.3$ (bulk)~\cite{Tanaka} \end{tabular} & \textbf{-3.99}~$\pm$~\textbf{0.04} & \begin{tabular}{@{}l@{}}\textbf{0.32}~$\pm$~\textbf{0.02}\\\textbf{0.64}~$\pm$~\textbf{0.02} (bulk)~~~\end{tabular} & \textbf{0.11}~$\pm$~\textbf{0.02} \\[0.5ex]
\hline
monolayer MoS$_{2}$ & \begin{tabular}{@{}l@{}}\textbf{-4.0}~$\pm$~\textbf{0.2}\\-4.6~$\pm$~0.08 (bulk)~\cite{Goto} \\ -3.1~$\pm ~ 0.1$ (bulk)~\cite{Tanaka} \end{tabular} & \textbf{-4.65}~$\pm$~\textbf{0.17} & -- & -- \\[0.5ex]
\hline
monolayer WSe$_{2}$ & \begin{tabular}{@{}l@{}}-3.7~$\pm$~0.2~\cite{Wang}\\-4.37~$\pm$~0.15 \cite{Srivastava}\\-1.57 to -2.86 \cite{Aivazian} \\ -4~$\pm$~0.5 \cite{Mitioglu}\end{tabular}& -- & -- & --\\
\hline
monolayer MoSe$_{2}$~~~~ & \begin{tabular}{@{}l@{}}-3.8~$\pm$~0.2~\cite{MacNeill,Wang}\\-4.1~$\pm$~0.2~\cite{Li}\end{tabular} & -- & -- & -- \\[0.5ex]
\hline

\end{tabular}
\caption{\textbf{Summary of effective g-factors and diamagnetic shifts in monolayer TMDs.} Experimentally measured exciton $g$-factors corresponding to the valley Zeeman effect, and diamagnetic shift coefficients $\sigma$, in monolayer transition-metal dichalcogenide materials. Measurements from this work are indicated in boldface. For comparison, also included are selected measurements from bulk crystals.}
\label{table:gfactors}
\end{table*}

\textbf{Non-local dielectric screening in monolayer TMDs.} In addition to the reduced mass $m_r$, the characteristic \emph{size} of the A and B excitons in monolayer TMDs is an essential parameter for determining material and optical properties. This is especially relevant because of \emph{non-local dielectric screening} in these and other 2D materials, which fundamentally modifies the functional form of the attractive potential $V(r)$ between electrons and holes \cite{Keldysh, Cudazzo, Berkelbach}. Rather than a conventional Coulomb potential, $V(r)$ is believed to assume the following form:
\begin{equation}
V(r)=-\frac{e^2}{8 \varepsilon_0 r_0}\left[ H_0 \left(\frac{r}{r_0} \right) - Y_0 \left(\frac{r}{r_0}\right) \right],
\end{equation}
where $H_0$ and $Y_0$ are the Struve function and Bessel function of the second kind, respectively, and the characteristic screening length $r_0 = 2 \pi \chi_{\rm 2D}$, where $\chi_{\rm 2D}$ is the 2D polarizability of the monolayer material \cite{Cudazzo, Berkelbach}. This potential follows a $1/r$ Coulomb-like potential for large electron-hole separations $r \gg r_0$, but diverges weakly as $\textrm{log}(r)$ for small separations $r \ll r_0$, leading to a markedly different Rydberg series of exciton states with modified wavefunctions and binding energies that cannot be described within a hydrogen-like model \cite{Berkelbach, Chernikov, He, Ye}.

\textbf{Diamagnetic shifts in monolayer WS$_2$.} To this end, the use of very large 65~T magnetic fields allows us to measure for the first time the small \emph{diamagnetic} shifts of excitons in monolayer TMDs so that the characteristic size of their wavefunctions can be directly inferred. In general \cite{Knox, Miura, Walck}, an exciton diamagnetic shift $\Delta E_{\rm dia}$ is expressed as
\begin{equation}
\Delta E_{\rm dia} = \frac{e^2}{8 m_r} \langle r^2 \rangle B^2 = \sigma B^2.
\end{equation}
Here, $\sigma$ is the diamagnetic shift coefficient, $m_r$ is the in-plane reduced mass, $r$ is a radial coordinate in a plane perpendicular to the applied magnetic field $B$ (here, for $B \parallel \hat{z}$, $r$ is in the monolayer plane), and $\langle r^2 \rangle_{1s} = \langle \psi_{1s}| (x^2 + y^2) |\psi_{1s} \rangle$ is the expectation value of $r^2$ over the 1$s$ exciton wavefunction $\psi_{1s}(r)$. Equation (2) applies in the `low-field' limit where the characteristic cyclotron energies $\hbar \omega_c$ (and also $\Delta E_{\rm dia}$) are less than the exciton binding energy, which is the case for excitons in TMDs even at $\pm$65~T. Given $m_r$, $\sigma$ can then be used to determine the rms radius of the 1$s$ exciton in the monolayer plane, $r_1$:
\begin{equation}
r_1 \equiv \sqrt{\langle r^2 \rangle_{\rm 1s}} = \sqrt{8 m_r \sigma}/e.
\end{equation}
This definition is entirely general and independent of $V(r)$. [Note that for a standard Coulomb potential $V(r) = -e^2/(4\pi \varepsilon_r \varepsilon_0 r$) in two dimensions, $r_1 = \sqrt{\frac{3}{2}}a_{0, \textrm{2D}}$, where $a_{0, \textrm{2D}} = 2\pi \varepsilon_r \varepsilon_0 \hbar^2/m_r e^2$ is the classic Bohr radius for hydrogenic 2D excitons.]

Exciton diamagnetic shifts have eluded detection in recent magneto-PL studies of monolayer MoSe$_2$ and WSe$_2$ \cite{Li, MacNeill, Srivastava, Aivazian, Wang}, likely due to the limited field range employed ($|B| < 10$~T). Here, the diamagnetic shift of the A exciton in monolayer WS$_2$ can be seen in 65~T fields via the slight positive curvature of both $E^+(B)$ and $E^-(B)$ in Fig. 2d. To directly reveal $\Delta E_{\rm dia}$, Fig. 2g shows the \emph{average} exciton energy, $(E^+ + E^-)/2$.  Overall quadratic shifts are indeed observed, indicating diamagnetic coefficients $\sigma_\textrm{A}= 0.32 \pm 0.02~\mu$eV/T$^2$ for the A exciton and a smaller value of $\sigma_\textrm{B} = 0.11 \pm 0.02~\mu$eV/T$^2$ for the B exciton. These measurements were repeated on five different regions of the monolayer WS$_2$ sample, with similar results.

\textbf{Exciton radii and binding energies.} Importantly, knowledge of $\sigma$ constrains not only the rms exciton radius $r_1$ (if the mass is known), but also the exciton binding energy if the potential $V(r)$ is known. Theoretical estimates \cite{Xiao, Berkelbach, Ram, Shi} for the A exciton reduced mass in monolayer WS$_2$ range from 0.15 to 0.22$m_0$, from which we can then directly calculate $r_{1, \textrm{A}}=1.48-1.79$~nm via Eq.~(3). These values are in reasonable agreement with recent \emph{ab initio} calculations of the 1$s$ exciton wavefunction in monolayer WS$_2$ \cite{Ye}, and further support a picture of 2D Wannier-type excitons with lateral extent larger than the monolayer thickness (0.6~nm) and spanning several in-plane lattice constants.

Moreover, $\sigma$, $m_r$, and $r_1$ can then be used to calculate the A exciton wavefunction $\psi_{1s}(r)$ and its binding energy, by numerically solving the 2D Schr\"odinger equation for describing the relative motion of electrons and holes using the potential $V(r)$ as defined in Eq.~(1), and taking the screening length $r_0$ as an adjustable parameter to converge on a solution for $\psi_{1s}(r)$ that has the correct rms radius $r_1$. For example, using $m_{r,\rm A}=0.16 m_0$ for the A exciton in WS$_2$, and using the measured diamagnetic shift $\sigma_\textrm{A}$, we find that $r_{1, \textrm{A}}= 1.53$~nm via Eq. (3). A wavefunction $\psi_{1s}(r)$ with this rms radius, shown explicitly in Fig.~3a, is calculated if (and only if) the screening length $r_0=5.3$~nm, and the binding energy of this state is 410~meV. For comparison, this inferred screening length is somewhat larger than expected for a suspended WS$_2$ monolayer (where $r_0 = 2 \pi \chi_{\rm 2D} = 3.8$~nm \cite{Berkelbach}), but is less that the value of 7.5~nm used recently by Chernikov \cite{Chernikov} to fit a non-hydrogenic Rydberg series of excitons in WS$_2$ from reflectivity data. Similarly, the 410~meV exciton binding energy that we estimate exceeds the value inferred by Chernikov (320~meV), but is less than the 700-830~meV binding energies extracted from two-photon excitation studies \cite{Ye, Zhu} and reflectivity/absorption studies \cite{Hanbicki} of monolayer WS$_2$. We emphasize, however, that the exciton wavefunctions and binding energies that we calculate necessarily depend on the reduced mass $m_r$ and the exact form of the potential $V(r)$, which is sensitive to the details of the dielectric environment and choice of substrate material \cite{Komsa, Lin}.

More generally, Fig.~3b shows a color-coded surface plot of the exciton binding energy, calculated within the framework of the non-local dielectric screening potential $V(r)$ defined in Eq.~(1), over a range of reduced masses $m_r$ and effective dielectric screening lengths $r_0$. At each point, the 1$s$ exciton wavefunction $\psi_{1s}(r)$, its binding energy, and its rms radius $r_1$ were calculated, from which we computed the expected diamagnetic shift coefficient $\sigma = e^2 r_1^2/8 m_r$. Importantly, the solid lines on the plot indicate the \emph{contours of constant diamagnetic shift} that correspond to our experimentally-measured values $\sigma_\textrm{A}$ and $\sigma_\textrm{B}$. At intervals along these contours, $r_1$ is indicated. From this plot, it can be immediately seen that over the range of theoretically-calculated masses ($m_{r, \textrm{A}}= 0.15-0.22 m_0$), excitons having the appropriate size to give the measured diamagnetic shift $\sigma_\textrm{A}$ (\emph{i.e.}, those lying along the $\sigma_\textrm{A}$ contour) have binding energies in the range of 480-260~meV.  Within this model, excitons with even larger binding energies (but still constrained to exhibit the correct diamagnetic shift) are anticipated if the reduced mass $m_r$ is lighter and the effective screening length $r_0$ is smaller.

\begin{figure}[tbp]
\center
\includegraphics[width=.48\textwidth]{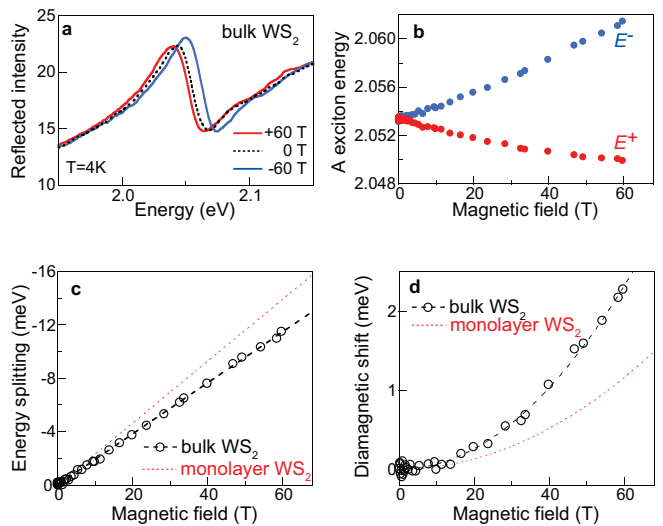}
\caption{\textbf{Zeeman splitting and diamagnetic shift in bulk WS$_2$.} (a) Intensity of reflected $\sigma^+$ light from the A exciton in bulk WS$_2$ at 4~K, using $B$=0, +60, and $-$60~T. (b) Energies of the Zeeman-split exciton transitions, $E^+$ and $E^-$. (c) The measured exciton splitting ($E^+ - E^-$), corresponding to $g=-3.33$. For comparison, the red line shows the valley Zeeman splitting measured in monolayer WS$_2$ (\emph{cf} Fig. 2f). (d) The average energy, ($E^+ + E^-$)/2, showing a diamagnetic shift (0.64~$\mu$eV/T$^2$) that is twice as large as that measured in monolayer WS$_2$ (red line; $\emph{cf}$ Fig. 2g).} \label{fig3}
\end{figure}

In addition, Fig.~3b also allows us to estimate the mass, binding energy, and spatial extent of the B exciton in monolayer WS$_2$, for which a smaller diamagnetic shift of $\sigma_\textrm{B}=0.11~\mu$eV/T$^2$ was measured.  Assuming that the local dielectric environment is similar for A and B excitons (\emph{i.e.}, $r_0$ is unchanged), then parameters for the B exciton lie at a point on the $\sigma_\textrm{B}$ contour that is directly to the right of those on the $\sigma_\textrm{A}$ contour. Thus, if $m_{r,\textrm{A}} = 0.16 m_0$ and $r_{1,\textrm{A}}=1.53$~nm as discussed above, then the B exciton reduced mass is $m_{r,\textrm{B}} = 0.27 m_0$, its rms radius is $r_{1,\textrm{B}}=1.16$~nm, and its binding energy is 470~meV. These values are qualitatively consistent with trends identified in early optical studies of bulk MoS$_2$ crystals \cite{Evans1968, Neville}, in which B exciton masses and binding energies were found to exceed those of A excitons.  These results highlight a further interesting consequence of the potential $V(r)$, which is that exciton binding energies scale only weakly and non-linearly with $m_r$, in contrast to the case for hydrogenic potentials.

\textbf{Zeeman splitting and diamagnetic shifts in bulk WS$_2$}. For comparison with monolayer WS$_2$, circularly-polarized magneto-reflectance measurements were also performed on a \emph{bulk} WS$_2$ crystal, where the A exciton resonance and its Zeeman splitting are readily resolved (Figs. 4a, b). The Zeeman splitting, shown in Fig. 4c,  increases linearly with field at a rate of $-193$~$\mu$eV/T ($g= -3.33$), in excellent agreement with early magnetic circular dichroism measurements of $g$-factors in bulk WS$_2$ \cite{Tanaka}, wherein it was suggested that deviations from $g=-4$ arise from the crystal-field mixing of $p$-type chalcogen atomic orbitals into the predominantly $d$-type character of the conduction and valence bands. Within this context, the value of $g \simeq -4$ that we measured in monolayer WS$_2$ (see Fig. 2f) suggests that such mixing effects may be suppressed in atomically-thin WS$_2$. In addition, Fig. 4d shows that the measured diamagnetic shift of the A exciton in bulk WS$_2$ is 0.64~$\mu$eV/T$^2$, which is twice as large as in monolayer WS$_2$.  Assuming an in-plane reduced mass of $m_r = 0.21 m_0$ in bulk WS$_2$ \cite{Beal1976}, we calculate via Eq. (3) an in-plane rms radius of 2.48~nm for the bulk A exciton, which is substantially larger than inferred for monolayer WS$_2$. Further, using an effective dielectric screening constant $\varepsilon_r = 7.0$ \cite{Beal1976}, we estimate the A exciton binding energy in bulk WS$_2$ via the standard hydrogenic formulation, $m_r/(m_0 \varepsilon_r^2) \times 13.6$~eV = 58~meV, which is close to that found in other bulk TMDs~\cite{Beal1972}. Therefore, we estimate that the binding energy of the A exciton in WS$_2$ increases by approximately a factor of 7 upon reducing the dimensionality of the host crystal from 3D to 2D. Note, however, that these estimates depend on the assumed value of the reduced mass $m_r$, which has not yet been measured independently by, \emph{e.g.}, cyclotron resonance studies.

\begin{figure}[tbp]
\center
\includegraphics[width=.48\textwidth]{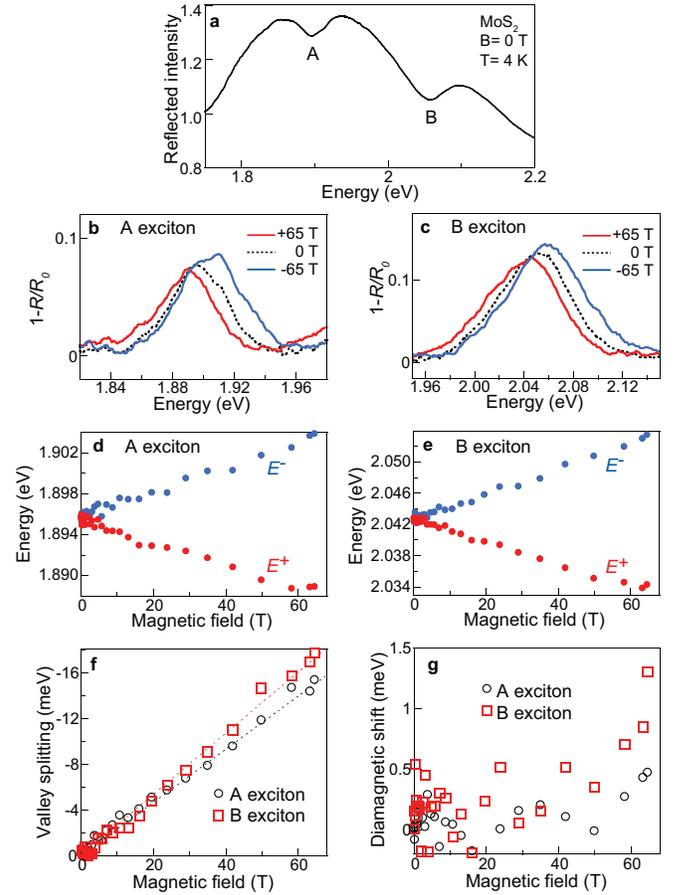}
\caption{\textbf{Valley Zeeman effect in monolayer MoS$_2$.} (a) Reflection spectrum of monolayer MoS$_2$ at $B$=0~T and $T$=4~K. The A and B exciton resonances are labeled. (b,c) Normalized reflection spectra (1-$R/R_0$) at the A and B exciton resonances at $B$=0, +65~T and -65~T. (d) Energies of the field-split A exciton transition. (e) Same, but for the B exciton transition. (f) The measured valley splitting ($E^+ - E^-$) versus magnetic field, for both A and B excitons. (g) The average energy of the two valley-split resonances, ($E^+ + E^-$)/2, for both A and B excitons.  Increased scatter in the data from this MoS$_2$ sample precludes any clear identification of the diamagnetic shift.} \label{fig4}
\end{figure}

\textbf{Valley Zeeman effect in monolayer MoS$_2$}. To complete this study of the monolayer transition-metal disulphides, we also performed high-field magneto-reflectance studies on large-area samples of monolayer MoS$_2$ (Fig.~5). The A and B exciton linewidths are broader and the optical reflection contrast is lower than for monolayer WS$_2$ (Fig. 5a). Nonetheless, a clear valley Zeeman splitting of both excitons is observed (Figs. 5b, c). The energies of the field-split exciton peaks are shown in Figs. 5d and 5e for the A and B excitons, respectively. Although the reduced signals and broader features lead to increased scatter in the fitted data, Fig.~5f shows that the measured valley splitting of the A and B excitons in MoS$_2$ increases approximately linearly with field at rates of $-233~\pm~10~\mu$eV/T and $-270~\pm~10~\mu$eV/T, corresponding to $g \simeq~-4.0~\pm 0.2$ and $-4.65~\pm~0.17$, respectively.  For the A exciton, this value is very close to those inferred from low-field magneto-PL studies \cite{MacNeill, Li, Wang} of their diselenide counterpart, monolayer MoSe$_2$ (see Table 1). As discussed above for the case of monolayer WS$_2$, a $g$-factor of $-4$ for the A exciton agrees surprisingly well with expectations from a simple two-band tight-binding picture, and suggests that the valley Zeeman effect in MoS$_2$, much like MoSe$_2$, is largely uninfluenced by contributions from the valley orbital (Berry curvature) magnetic moment. We note, however, that the measured valley $g$-factor is somewhat larger for the B exciton in monolayer MoS$_2$. Unfortunately, the reduced signal levels from these monolayer MoS$_2$ samples led to correspondingly increased scatter in the fitted exciton energies, precluding an accurate determination of exciton diamagnetic shifts in monolayer MoS$_2$ (see Fig. 5g).

In summary, we have presented a comprehensive study of valley Zeeman effect and diamagnetic shifts of excitons in the archetypal monolayer transition metal disulphides WS$_2$ and MoS$_2$. Valley $g$-factors of the A excitons are approximately $-4$, which are similar to those obtained from transition metal diselenides.  Unexpectedly, the heavier B exciton in monolayer WS$_2$ also exhibits $g \simeq -4$, suggesting that the valley Zeeman effect is largely unaffected by the exciton reduced mass. The very large magnetic fields used in these studies also allowed the first measurements of the exciton diamagnetic shifts in a monolayer TMD -- specifically, WS$_2$ -- from which rms exciton radii were directly computed ($r_1 = 1.53$~nm and 1.16~nm for the A and B excitons, respectively). Within a picture of non-local dielectric screening in these 2D semiconductors, these measurements of diamagnetic shifts allowed us to constrain estimates of the exciton binding energies, which we calculate to be 410~meV and 470~meV for the A and B excitons in monolayer WS$_2$. These studies highlight the utility of very large magnetic fields for characterizing new 2D material systems.

We thank K. Velizhanin for helpful discussions.  These optical studies were performed at the National High Magnetic Field Laboratory, which is supported by NSF DMR-1157490 and the State of Florida. Work at NRL was supported by core programs and the NRL Nanoscience Institute, and by AFOSR under contract number AOARD 14IOA018-134141. J. K. was supported by the Air Force Office of Scientific Research under Award Number FA9550-14-1-0268.

\end{document}